\documentstyle[12pt]{article}
\input epsf

\begin{document}
\renewcommand{\thefootnote}{\fnsymbol{footnote}}
\begin{titlepage}
\renewcommand{\thefootnote}{\fnsymbol{footnote}}
\makebox[2cm]{}\\[-1in]
\begin{flushright}
\begin{tabular}{l}
TUM/T39-96-21
\end{tabular}
\end{flushright}
\vskip0.4cm
\begin{center}
  {\Large\bf Combining Lattice QCD Results with Regge Phenomenology in a
    Description of Quark Distribution Functions\footnote{Work 
    supported in part by BMBF}}\\
\vspace{2cm} 
T. Weigl and L.Mankiewicz\footnote{On leave of absence from N. Copernicus
Astronomical Center, Polish Academy of Science, ul. Bartycka 18,
PL--00-716 Warsaw (Poland)}\\
\vspace{1.5cm}
{\em Institut f\"ur Theoretische Physik, TU M\"unchen, Germany}\\
\vspace{1cm}
{\em \today}\\
\vspace{1cm} 
{\bf Abstract:\\[5pt]} \parbox[t]{\textwidth}{
  The most striking feature of quark distribution functions
  transformed to the
  longitudinal distance representation is the recognizable separation of small
  and large longitudinal distances. While the former are responsible for the
  average properties of parton distributions, the latter can be shown to
  determine specifically their small-$x$ behavior. In this paper we
  demonstrate how the distribution at intermediate longitudinal distances can
  be approximated by taking into account constraints which follow from the
  general properties of parton densities, such as their support and behavior
  at $x \to 1$.  We show that the combined description of small, intermediate,
  and large longitudinal distances allows a good 
  approximation of both shape
  and magnitude of parton distribution functions.  As an application we have
  calculated low-virtuality C even and odd (valence) u and d quark
  parton densities of the
  nucleon and the C-even transversity distribution $h_1(x)$, combining
  recent QCD sum rules and lattice QCD results with phenomenological
  information about their small-$x$ behavior.}

\vspace{1cm}
\end{center}
\end{titlepage}
\newpage

More than twenty five years ago the discovery of point-like constituents,
partons, in the deep-inelastic-scattering revolutionized our
understanding of the nucleon structure\cite{partons}. Although since
then the Quantum Chromodynamics (QCD) has become an established theory
of strong interactions \cite{QCD89}, derivation of parton distribution
functions from the first principles still remains a challenge. The
standard QCD analysis based on the Operator Product Expansion (OPE)
relates moments of parton distributions at a given scale to the
nucleon matrix elements of local twist-2 operators \cite{OPE}. The
evolution of parton distributions, due to the renormalisation scale
dependence of twist-2 quark and gluon operators, is completely
understood within QCD perturbation theory \cite{Blum95}.  Hence, the
whole problem can be reduced to the sufficiently accurate,
first-principles computation of parton distributions at low
virtuality, which in practice has proven to be quite difficult. For
example, QCD sum rules calculations of parton distributions have been
only moderately successful \cite{SumRules} - the results for 
moments beyond the lowest non-trivial one have typically very low accuracy
\cite{Ross96} - and the reliable, state-of-art lattice QCD predictions for
the lowest two moments of C-even quark distribution functions have
become available only recently \cite{LAT95}.


For more than twenty years, the intrinsic complication of QCD has prevented us
from obtaining its exact solutions in the non-perturbative domain. This
explains the important role played by phenomenological investigations and
indeed since the discovery of partons there have been many attempts to model
parton distribution functions, for example in terms of non-relativistic
potential models \cite{NRP}, relativistic bag models \cite{BAG}, or soliton
models \cite{SOL}.  Methodologically, all these calculations can be understood
as some effective low-energy theories of hadronic structure. All of them have
either semi-classical or quantum mechanical character and they do not correctly
take into account mechanisms specific to a quantum field theory, such as
e.g. the 
existence of quarks and gluon loops.  Moreover, despite the current wisdom that
quarks are the most natural effective degrees of freedom in this context, one
can argue from the very beginning that
there might be a problem with such a choice in the case of
low virtuality parton distribution functions. Certainly, the SU(6) algebraic
properties of the hadronic spectrum suggest that quarks are very useful
effective degrees of freedom as far as static hadronic properties, such as
masses or electroweak parameters are concerned, 
but the same does not have to be true for structure
functions, which are related to the {\it infinite} series of twist-2 matrix
elements. 
On the other hand, quark models as an effective low-energy description cannot
be considered capable to reproduce all of them with a reasonable accuracy.
The variety of successful models of hadronic static
properties shows that the latter are not very sensitive to details of
phenomenological approximation, and
therefore can be well understood with the help of an effective, low-energy
description. In the case of structure functions the situation is less clear. It
is not easy to construct a model which satisfactory reproduces the overall
magnitude and the $x$-dependence of flavor-singlet structure functions without
having to incorporate some additional assumptions which sometimes are difficult
to justify. For example, one starts with a calculation which produces
a valence - like distribution which is subsequently perturbatively evolved to
higher scales, where DIS measurements are performed. Unfortunately, consistency
with the data often requires that the starting point of the evolution has to be
very low, deeply in the non-perturbative domain where the perturbative approach
to the evolution is certainly controversial.

One very practical aspect of this situation is that the input parton
distributions for modern parametrizations of structure functions
\cite{MRS,CTEQ,GRV94} do not yet have a satisfactory i.e., fundamental,
physical interpretation. Typically a simple functional form is evolved to
higher $Q^2$ and fitted to the data in this domain.  This program has been
applied so successfully in practice that the $Q^2$ dependence of structure
functions, governed by QCD evolution equations, has become the key factor for
our understanding of experimental data, and the major argument which strongly
points to QCD as the theory of strong interactions \cite{Blum95}. To the
contrary, the magnitude and $x$-dependence of the deep-inelastic cross-section
still belong to the much less understood, non-perturbative aspects. In
principle, the OPE supplemented by non-perturbative methods such as QCD sum
rules, models of the QCD vacuum \cite{Shur96}, or ultimately the lattice
calculations should allow to predict the cross-section as a whole, but as we
have already mentioned, in practice this problem is very difficult.

In a recent paper we have attempted to optimize the applicability of OPE to the
problem of hadronic structure functions $q(x,\mu^2)$ \cite{Weig96}. The key
observation is that even at the twist-2 level their Bjorken $x$-dependence
reflects physics of two very different regimes, which can be identified and
separated by the following transformation into the coordinate space,
\begin{eqnarray}
Q(z,\mu^2) & = & \int_0^1 du \, q(u,\mu^2) \sin{u z} \, , \nonumber \\
q(x,\mu^2) & = & \frac{2}{\pi} \int_0^\infty dz \, Q(z,\mu^2) \sin{u z}\, ,
\label{quarkCplus}
\end{eqnarray}
where the distribution function $Q(z,\mu^2)$ has an interpretation of the
gauge-invariant correlation function of two quark fields at the light cone,
i.e. in QCD it is defined by the nucleon matrix element of a non-local quark
string operator \cite{Col82,Bal88}. The Ioffe-time $z$, the coordinate variable
dual to the Bjorken $x$, is an invariant measure of the longitudinal distance
along the light-cone between the quark fields \cite{Iof65,Duc92}.  The Taylor
expansion of $Q(z,\mu^2)$ around the origin has coefficients given by moments
of the corresponding momentum space parton distribution, or equivalently by the
matrix elements of twist-2 local operators \cite{Col82}. As in the following we
are interested in parton distributions at low virtualities, we choose a
normalization point $\mu^2$ to be as low as possible from the common sense
point of view, somewhere between 1 and, say, 3-4 GeV$^2$.

As it has been found in \cite{Weig96,Bra95,Hoy96}, the experimental data shows
completely different behavior of Ioffe-time distributions for small and large
longitudinal distances, see Figures 1 and 2 of Ref.\cite{Weig96}. For small
$z$, the 
distribution is almost linear, which means that it can be described by the
first one or two terms of its Taylor expansion around the origin, given by the
matrix elements of the two local twist-2 operators of the lowest dimension.
When the variable $z$ reaches $ \sim 5$, the character of the distribution
changes and it turns to a almost flat, asymptotic form 
for $z \ge 10$. In
this region the shape of $Q(z)$ is determined by the small-$x$ behavior of the
corresponding parton densities:
\begin{equation}
q(x) \sim \frac{1}{x^{1+\alpha}} \Longleftrightarrow Q(z) \sim z^{\alpha}\, .
\end{equation}
Note that $z = 10$ corresponds in the nucleon rest frame to the linear distance
of the order of 2 fm. In this region the contribution from valence partons
i.e., 
partons carrying the momentum fraction $x \ge 0.1$ becomes subleading in
comparison with that of wee partons from the domain $x \leq 0.1$ \cite{Hoy96}.
For $z \ge 10$ the corresponding quark-quark correlation function involves
fields separated by a distance larger then the electromagnetic nucleon size.
Naively, it is tempting to assume that the correlation function should vanish
in this region, to the contrary to what can be seen on Figures 1 and 2 of Ref.\cite{Weig96}, where
the distribution stays approximately constant, or even rises at large $z$. This
behavior is usually explained by the contribution of the virtual photon
transition into a quark-antiquark pair which occurs "outside" the nucleon, but
one has to realize that such an interpretation assumes that the target nucleon
indeed occupies a well defined space-time volume. We stress that the validity
of this assumption depends on the relevant quantum numbers in the process under
investigation. In particular, the finiteness of the hadronic 
electromagnetic radius
shows that electromagnetic charge, a C-odd quantity, is distributed in a finite
volume associated with the hadronic state. In a C-even process such as the
Compton scattering this corresponds to a transition between the valence and wee
parton regimes around longitudinal distance of 2 fm in the nucleon rest frame.
Qualitatively, degrees of freedom responsible for large-$z$ behavior of $Q(z)$
give no important contribution to, say, nucleon electromagnetic formfactor.
This observation can explain successful application of quark models to
charge-parity odd quantities like charge radii and magnetic moments and their
difficulties with parton distribution functions.

The very particular shape of the longitudinal distance distribution suggests a
new approach to the problem of reconstruction of parton distributions from
available theoretical information. We have already demonstrated \cite{Weig96}
that at small $z$ the parton distributions can be very well described by their
first two moments, alias calculable matrix elements of local twist-2 operators.
However, the expansion in terms of moments becomes less and less efficient when
$z$ approaches, and passes the transition region towards the large-$z$ domain.
Using the Pade approximation instead of Taylor expansion still requires three,
four moments to reach the onset of the asymptotic domain. Note that already the
third moment is related to the twist-2 operator with five covariant
derivatives, and it is probably fair to say that a reliable computation of its
matrix element using present day lattice technology would be very difficult. In
fact all the QCD methods available at present are based on the short
longitudinal distance expansion and therefore are not well suited to study the
large-$z$ region. In our opinion it is related to the fact that physics of
large and small longitudinal distances is different and it should be treated
separately.  In the previous paper we showed that one can combine the short
distance expansion and the assumption about large-$z$ alias small-$x$ behavior
to obtain the zeroth-order approximation showed as the dotted lines on Figures
1 and 2 \footnote{Please note that the corresponding curve depicted by the 
dotted line on Figure 5 in Ref.\cite{Weig96} is wrong due to 
an error in our computer
program.}.  There, $Q(z)$ has been approximated by two straight lines,
representing 
the small-$z$ and large $z$ behavior. Note that such a method clearly ignores
the information about the $z$ dependence in the transition region, and that it
results in the distribution functions which are not far from the data in the
small and intermediate $x$ domain, but become negative for large $x$. It is
therefore natural to assume that, given the known behavior of $Q(z)$ in the
small and large $z$ domains, the transition region could be constrained by the
requirement of the correct behavior of the parton distribution at $x \to 1$.
In fact, we have found it difficult to implement directly the positivity
constraint on the Bjorken-$x$ distributions in the Ioffe-time representation.
Instead, we imposed conditions that the distribution function vanishes at $x =
1$, together with at least its two derivatives, in agreement with the QCD
counting rules \cite{Bro95} which state that parton distributions
should vanish at least as $(1-x)^3$ at $x \to 1$.  The most obvious idea to use
the 
anticipated large-$x$ behavior to implement the subleading behavior of $Q(z)$
in the large and intermediate $z$ region turned out to be insufficient.
Instead, we have chosen to approximate the shape of the Ioffe-time distribution
in the small- and intermediate-$z$ region up to $z_a$ - the onset of the
asymptotic large-$z$ behavior, or 
the nucleon boundary, using the formula motivated by the Taylor expansion of
$Q(z)$, i.e.:
\begin{equation}
W(z) = a_1 z - a_3 z^3 + a_5 z^5 - a_7 z^7 + a_9 z^9 - a_{11} z^{11} \, . 
\label{poly}
\end{equation}
with the coefficients $a_1$ and $a_3$ given by the first two non-trivial
moments of the corresponding quark distribution function. We have found that
this is the only expansion which can be equally well applied to all three
different cases which we consider in this paper. Slightly better results can be
obtained when one uses in each case a method which anticipates the shape of the
Ioffe-time distribution one wants to reconstruct. Having the point $z_a
= 10$ fixed by the classical nucleon diameter, the 
continuity of $Q(z)$  eliminates one from four
free
parameters of (\ref{poly}), and the other three are determined using three
constraints imposed by the expected 
behavior at $x \to 1$. Altogether the whole procedure is equivalent to the
solution of a system of linear equations which can be easily found with the 
help of Mathematica. The results are plotted as dashed lines on Figures 1
and 
2 in comparison with solid lines which represent the MRS(A) parametrizations
\cite{MRS}. The improved approximation, despite its seemingly "kinematical"
character, is indeed quite good as far as both shapes and magnitudes of parton
distributions are concerned.  Let us summarize below once more all the
information which has been required to obtain this result:
\begin{itemize}
\item 
  the first two moments, which have been already computed on the
  lattice, at least in the quenched approximation,
\item 
  the shape of the longitudinal distance distribution at large-$z$, which is
  directly related to the small-$x$ behavior of the corresponding parton
  distribution function,
\item 
  the onset of the large longitudinal distance regime which, as we have
  discussed above, should in principle correspond to the nucleon
  electromagnetic size. Nevertheless, it is somewhat surprising that the data
  indeed show the onset of the asymptotics at $z$ 
  $\approx$ 2 fm in the nucleon rest frame,
\item 
  behavior of parton distribution at $x \to 1$ which follows from support
  properties and perturbative QCD counting rules.  In the longitudinal distance
  representation it allows to construct a solution satisfying simultaneously
  the positivity requirement, for which we have not been able to find an
  equivalent condition in terms of $Q(z)$.
\end{itemize}
The only required piece of information which does not have yet a transparent
physical interpretation is normalization at small-$x$. As we have already
argued in \cite{Weig96}, this information is rather difficult to extract from
the standard OPE analysis.  In this sense as long as a procedure aimed to
determine this missing element has not been proposed, our program cannot be
considered to be complete, and we have to rely on phenomenological information.
On the other hand it is tempting to suggest that the correct QCD description of
parton distribution functions has to rely on our understanding of large $z$
physics as well, and therefore to complete the whole program it is necessary to
develop a theoretical framework in which the required information is calculable
through a relation to, say, matrix element of a certain, possibly non-local,
QCD operator. The first step in this direction has been made recently in
\cite{Bal95}.

To illustrate the above considerations we have attempted to reconstruct the u
and d C-even parton distributions using the recent lattice results for their
first two moments \cite{LAT95}.  They correspond to the normalization point of
about 2 GeV$^2$ in the ${\overline{\rm MS}}$ scheme. To predict the
corresponding parton distributions we have to combine them with the
normalization of small-$x$ data at the same low scale. This information has
been extracted from the CTEQ NLO \cite{CTEQ} parametrization which has the low
starting point at 2.5 GeV$^2$.  The difference between LO and NLO parton
distributions is in our case irrelevant as it is much smaller than the expected
accuracy. The results are compared to the CTEQ parametrizations on Figures 3
and 4. 
Clearly, the lattice u-quark data overestimate true magnitude of the first two
moments. As a consequence, the predicted parton distributions is larger than
the real one at intermediate and large values of $x$. The agreement with the
d-quark distribution, even if perhaps a bit fortuitous, is certainly
impressive.  

The same method can be applied to reconstruct the valence quark distributions
\begin{eqnarray}
Q_{\rm val}(z,\mu^2) & = & \int_0^1 du \, q_{\rm val}(u,\mu^2) \cos{u z} \, , 
\nonumber \\
q_{\rm val}(x,\mu^2) & = & \frac{2}{\pi} \int_0^\infty dz \, Q_{\rm val}(z,\mu^2)
\cos{u z}\, ,
\label{val}
\end{eqnarray}
from their first moments and large-$z$/small-$x$ asymptotics. As it follows
from (\ref{val}), the value of $Q_{\rm val}(z=0)$ is given by the number of
valence quarks.  At large $z$ the Ioffe time distribution should fall as
$z^{-\alpha}$, where, according to the classical Regge theory \cite{Regge},
$\alpha \approx -0.5$. Hence, the C-odd correlation function of 
quark fields along the
light-cone slowly vanishes at large longitudinal distances. As in the
case of C-even parton distributions, as long as both the asymptotic behavior
at large $z$ and the first non-trivial moment of $q_{\rm val}$ are known, the
corresponding Ioffe time distribution can be interpolated between these two
regions. 
In this case the interpolating formula reads
\begin{equation}
W(z) = a_0 - a_2 z^2 + a_4 z^4 - a_6 z^6 + a_8 z^8 - a_{10} z^{10}  \, ,
\label{poly1}
\end{equation}
with $a_0$ given by the number of valence quarks and $a_2$ known from a
non-perturbative calculation of the first
non-trivial moment of C - odd quark distribution function, gives much
better results. Combining the recent lattice \cite{LAT95} results with
the small-$x$ normalization extracted from the 
CTEQ parametrization\cite{CTEQ} at $\mu^2 = 2.5$ GeV$^2$ we have obtained
results presented on Figures 5 and 6 \footnote{Extrapolating from the large $z$
domain, i.e., using the expansion in the inverse powers of $z^{-0.5}$ it is
possible to obtain
qualitatively very similar results using the QCD sum rules formula for the
u-quark Ioffe-time time distribution from Ref.\cite{Bra95}}. All the
coefficients $a_4 \dots a_{10}$ 
come out positive as expected. Due to the fact that the lattice results
for the u-quark are much closer to the experimental data, the
agreement with the parametrization is better in this case then for the C-even
distribution. One can try to improve the d-quark approximation
by requiring that more than two derivatives vanish at $x = 0$, however sooner
or later the interpolation becomes unstable
when the order of the corresponding polynomial is too high.

The above procedure can also be applied to the
proton transversity distribution $h_1^p(x,\mu^2)$ \cite{JaffeJi91}. 
This idea is especially attractive if 
one considers that the calculation of, say, the first two moments of
$h_1(x)$ is probably possible using the present lattice resources
\cite{Aoki96}, so this
structure function can be predicted before it will be measured in the next
round of polarized DIS experiments \cite{exp}! As
the lattice QCD predictions are not yet available, we have reinterpreted,
following the
discussion of \cite{Bra95}, the recent QCD sum
rules calculation \cite{Ioff94} of the $x$-dependence of $h_1^p(x)$ as
the calculation of the Taylor expansion of the Ioffe-time
distribution. According to Ref.\cite{Ioff94} the d-quark contribution is small,
so the proton and neutron transversity distributions are approximately
equal to 4/9 and 1/9 $h_1^u(x)$, respectively. Using results
of \cite{Ioff94} it is not difficult to extract corresponding QCD sum 
rules predictions for the first moment of the C-even u-quark transversity
distribution, given by the reduced matrix
element of the QCD local operator ${\bar u}(0)\gamma_5
\sigma^{+\perp}iD^+u(0)$. The result is finite and free of any infrared
singularities:
\begin{equation} 
\int_0^1 du \, u~h_1^u(u) \approx 0.3 \nonumber \\ 
\label{h1}
\end{equation}
Note that this estimate corresponds to a low virtuality of the order of a few
GeV$^2$.  Note also that the singular behavior of C-odd tensor charge $\int_0^1
du \, h_1^u(u)$, equal to the reduced matrix element of the operator ${\bar
u}(0) \gamma_5 \sigma^{+\perp}u(0)$, present in the OPE of Ref.\cite{Ioff94},
is related to the prohibited contributions from large t-channel
distances. Consequently, it will disappear when bilocal
power corrections \cite{BPC82} are properly taken into account. One striking
feature of the value of the QCD sum rules result (\ref{h1}) is its magnitude --
approximately equal to the magnitude of the first moment of the unpolarized
distribution $\int_0^1 dx x u(x,\mu^2)$ at the scale around 2 GeV$^2$. On the
other hand unitarity requires that $h_1^u(x) \leq u(x)$. Now,
because at small $x$ Regge arguments suggest that $h_1^u(x) \ll u(x)$, one can
suspect that the resulting
transversity distribution, reconstructed subject to the constraint (\ref{h1}),
will violate unitarity at intermediate and large values of $x$. To determine
the behavior at large $z$ we followed Regge
arguments \cite{Ioff94,Efre94} which suggest that the ratio of $h_1(x)/g_1(x)$
should be constant at small $x$ and a low normalization scale.  Matching the
Taylor expansion at small $z$, with the first coefficient given by (\ref{h1}),
to the 
large $z$ Regge asymptotics $\sim z^{-1.3}$, with the magnitude fixed simply
by the $g_1^p$ data at small $x$, 
we have obtained a prediction for $h_1^p(x,\mu^2)$ at low $\mu^2$
depicted as the dashed line on Figure 7. As expected, at large and
intermediate $x$ 
it is larger than the unpolarized distribution represented on Figure 7 by the
the solid line. Assuming that the result (\ref{h1})
overestimates the true value by $50\%$ one
obtains the dot-dashed line on Figure 7. This prediction is similar in
magnitude to the bag-model result of \cite{JaJi92}. 
One should also keep in mind that the complicated shape inherent to a C-even
Ioffe-time transversity distribution, which is equal to zero at $z = 0$, goes
over a maximum 
at $z \sim 4-5$, and vanishes again for large $z$, makes any approximation
procedure in principle more sensitive to yet unknown higher moments. 

Finally we point out once more that an attempt to "squeeze out" the most
relevant information 
from structure functions can serve as a basis for construction of an effective
theory of parton distributions. For example, if only the first two moments are
required, one can hope that they can be computed with a reasonable accuracy in
low-energy phenomenological approximations to QCD such as quark models or
soliton models, with parameters unambiguously determined by comparison with
other low-energy observables.  Also from the lattice QCD point of view the
precise calculation of the first one or two moments although technically very
difficult, is perfectly feasible using the state-of-art technology. The
small-$x$ dynamics is much more specific and indeed it has been difficult to
describe this region in the framework of a standard low-energy
phenomenology. 
One reason for these difficulties is certainly related to the fact that,
looking from the point of view of the longitudinal distance expansion, only
quantities like low moments of parton distribution functions which correspond
to short-distances can be understood by semi-classical or quantum
mechanical models of hadronic structure. Analogous description of the physics
of longitudinal distances larger than the nucleon electromagnetic size is much
less justified because in this region parton correlation functions are much
more sensitive to a proper treatment of these aspects of QCD dynamics which are
difficult to represent using effective quark degrees of freedom. In QCD
the required 
information about the normalization of small-$x$ distributions can be perhaps
related \cite{Bal95}
to matrix elements of some non-local QCD operators. In this language our
discussion can be interpreted as an indication that in fact the required
smearing does 
not have to extend over longitudinal distances larger than $\sim 2$ fm in the
nucleon rest frame. 

\vspace{1cm} 

{\bf Acknowledgments} This work was supported
in part by BMBF and by KBN grant 2~P03B~065~10. LM gratefully acknowledges
discussions with S.J. Brodsky, X. Ji, S. Liutti, and T. Goldman during his
visit to the Institute for Nuclear Theory at the University of Washington.

\clearpage

\epsfbox{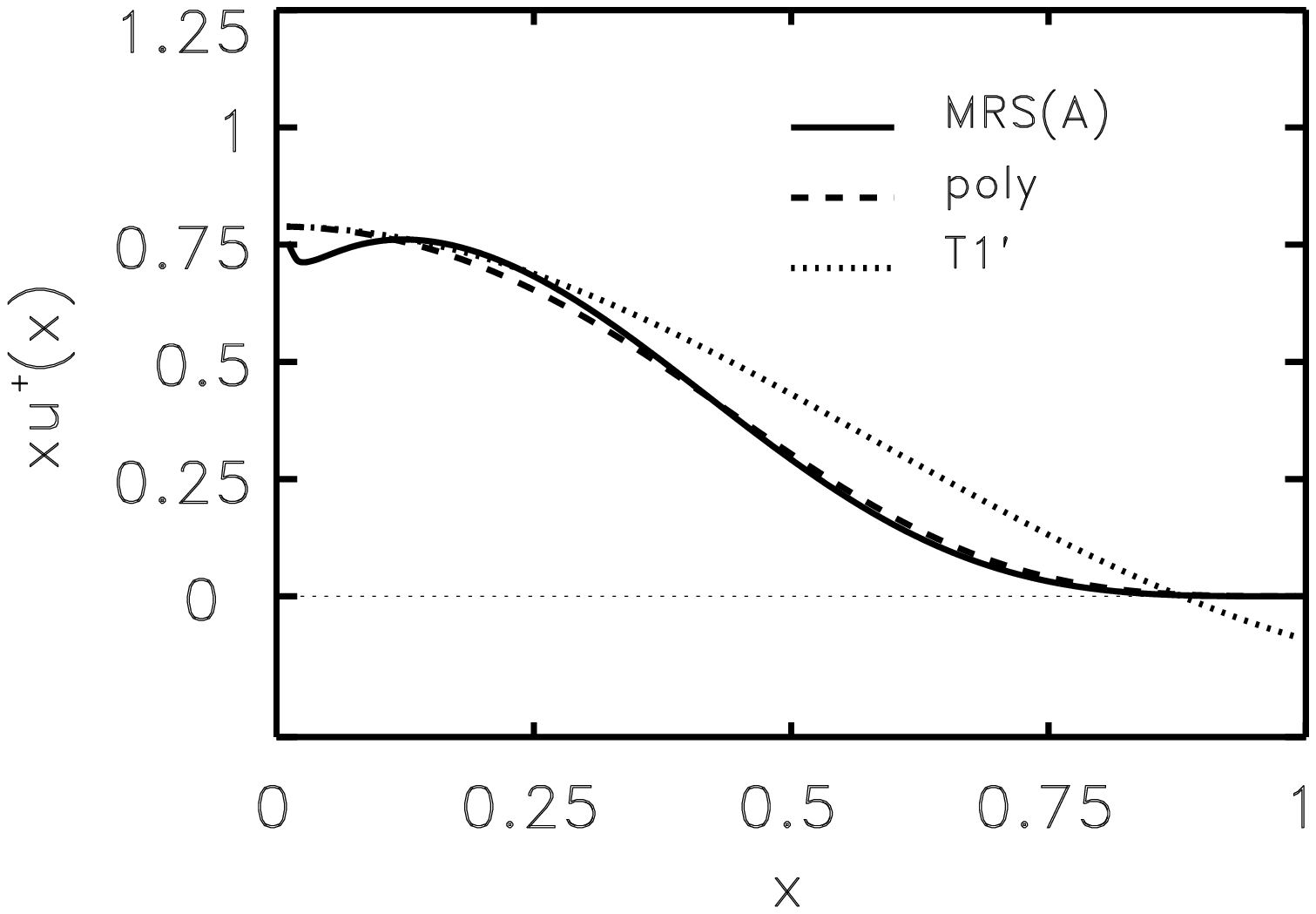}
\begin{description}
  
\item[Fig.~1] Consecutive approximations of the u-quark parton distribution
  $u(x)$, solid line, calculated using the MRS(A) parametrization at $\mu^2 = $
  4 GeV$^2$. The first approximation (dotted line) considered in \cite{Weig96}
  ignores the existence of the transition region between small and large
  Ioffe-time regimes. The improved approximation (dashed line) is obtained with
  the help of an interpolation between these two domains, subject to
  constraints on the behavior of Bjorken $x$ distribution at $x \to 1$.

\end{description}

\epsfbox{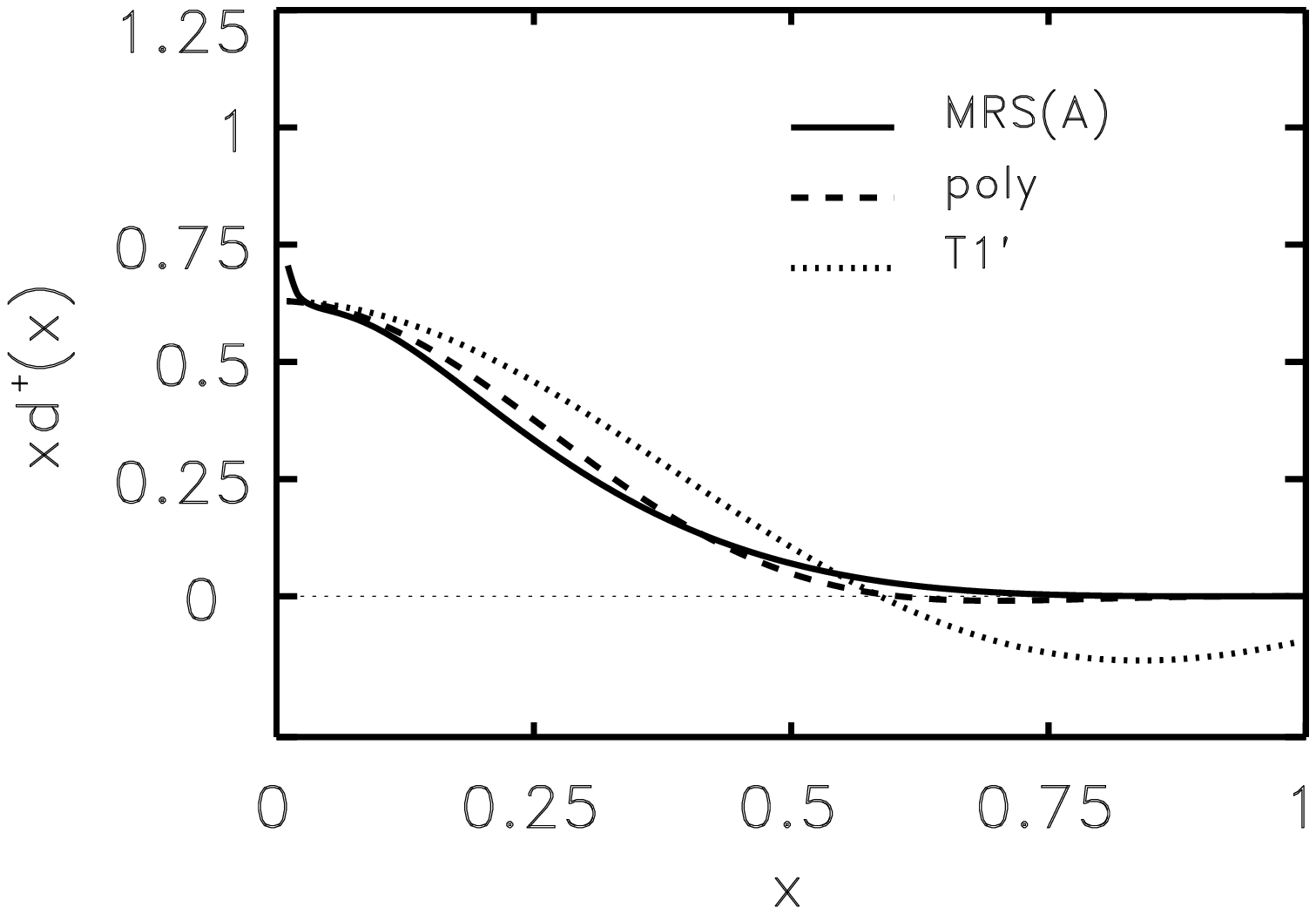}
\begin{description}
  
\item[Fig.~2] Consecutive approximations of the d-quark parton distribution
  $d(x)$, solid line, calculated using the MRS(A) parametrization at $\mu^2 = $
  4 GeV$^2$. The first approximation (dotted line) considered in \cite{Weig96}
  ignores the existence of the transition region between small and large
  Ioffe-time regimes. The improved approximation (dashed line) is obtained with
  the help of an interpolation between these two domains, subject to
  constraints on the behavior of Bjorken $x$ distribution at $x \to 1$.

\end{description}

\epsfbox{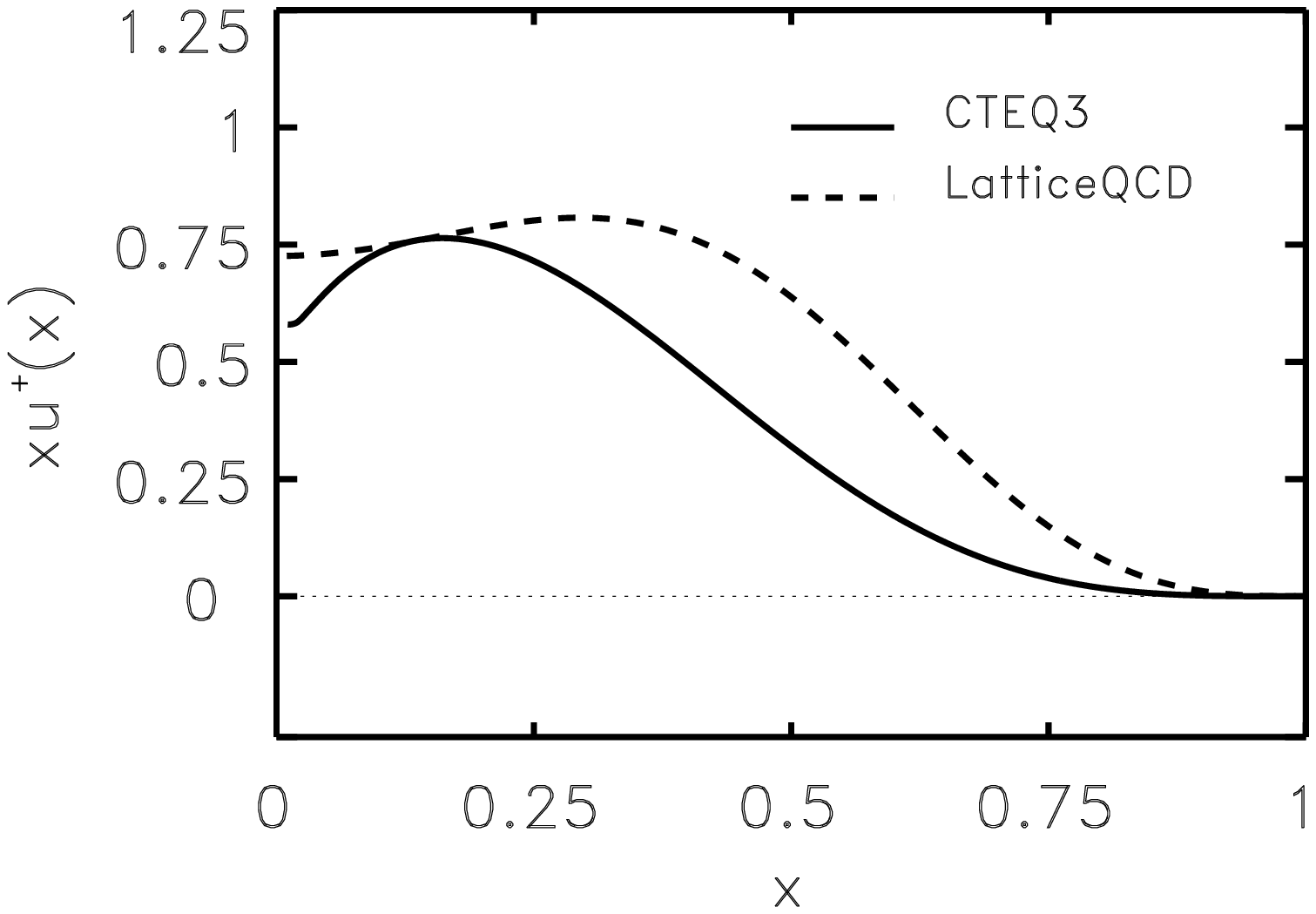}
\begin{description}
  
\item[Fig.~3] Reconstruction of the u-quark distribution function using the
  lattice results for the first two moments \cite{LAT95} and the experimental
  normalization at small $x$. The solid line is the CTEQ parametrization
  \cite{CTEQ} at $\mu^2 = 2.5$ GeV$^2$.

\end{description}

\epsfbox{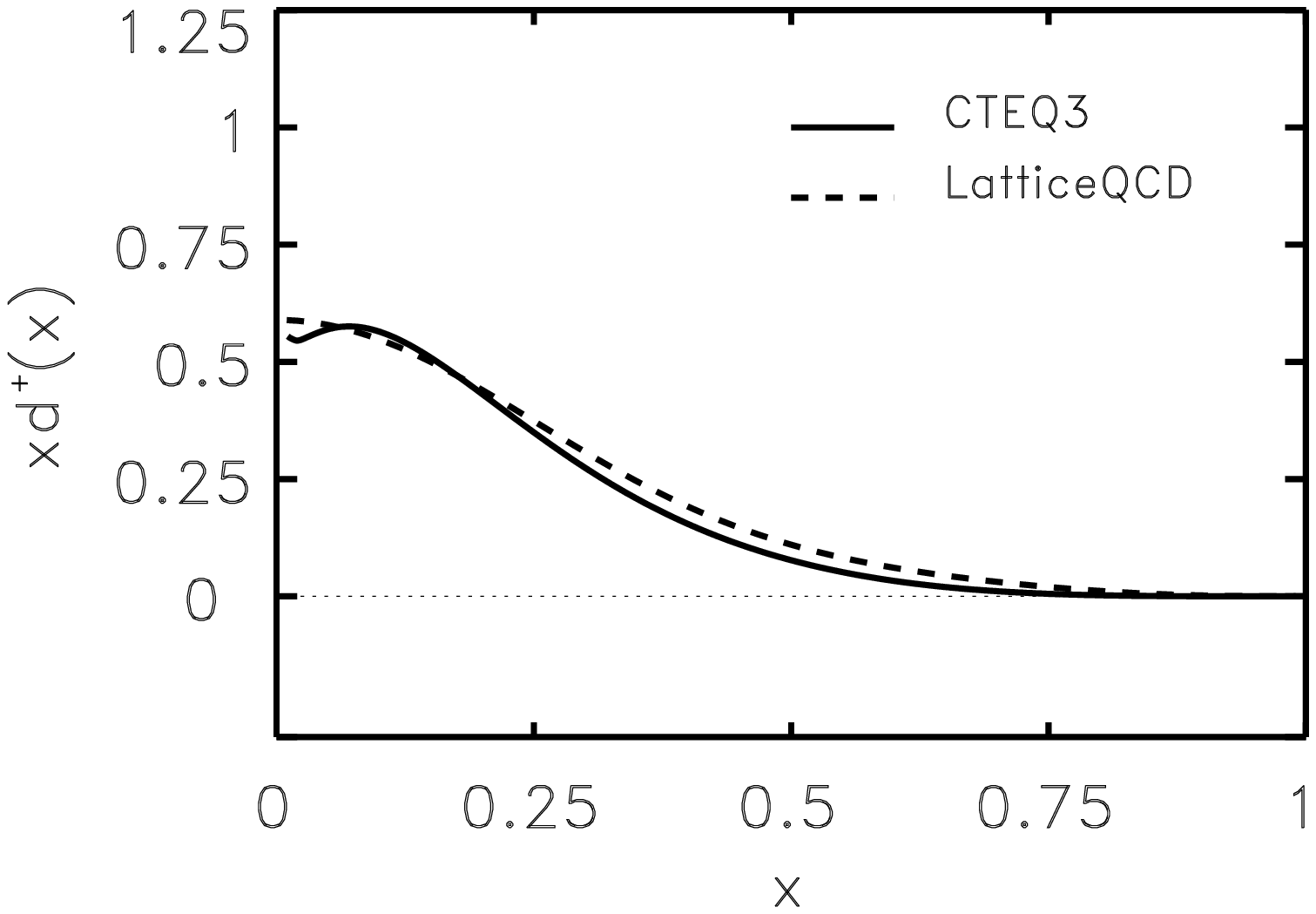}
\begin{description}
  
\item[Fig.~4] Reconstruction of the d-quark distribution function, dashed line,
  using the
  lattice results for the first two moments \cite{LAT95} and the experimental
  normalization at small $x$. The solid line is the CTEQ parametrization
  \cite{CTEQ} at $\mu^2 = 2.5$ GeV$^2$.

\end{description}

\epsfbox{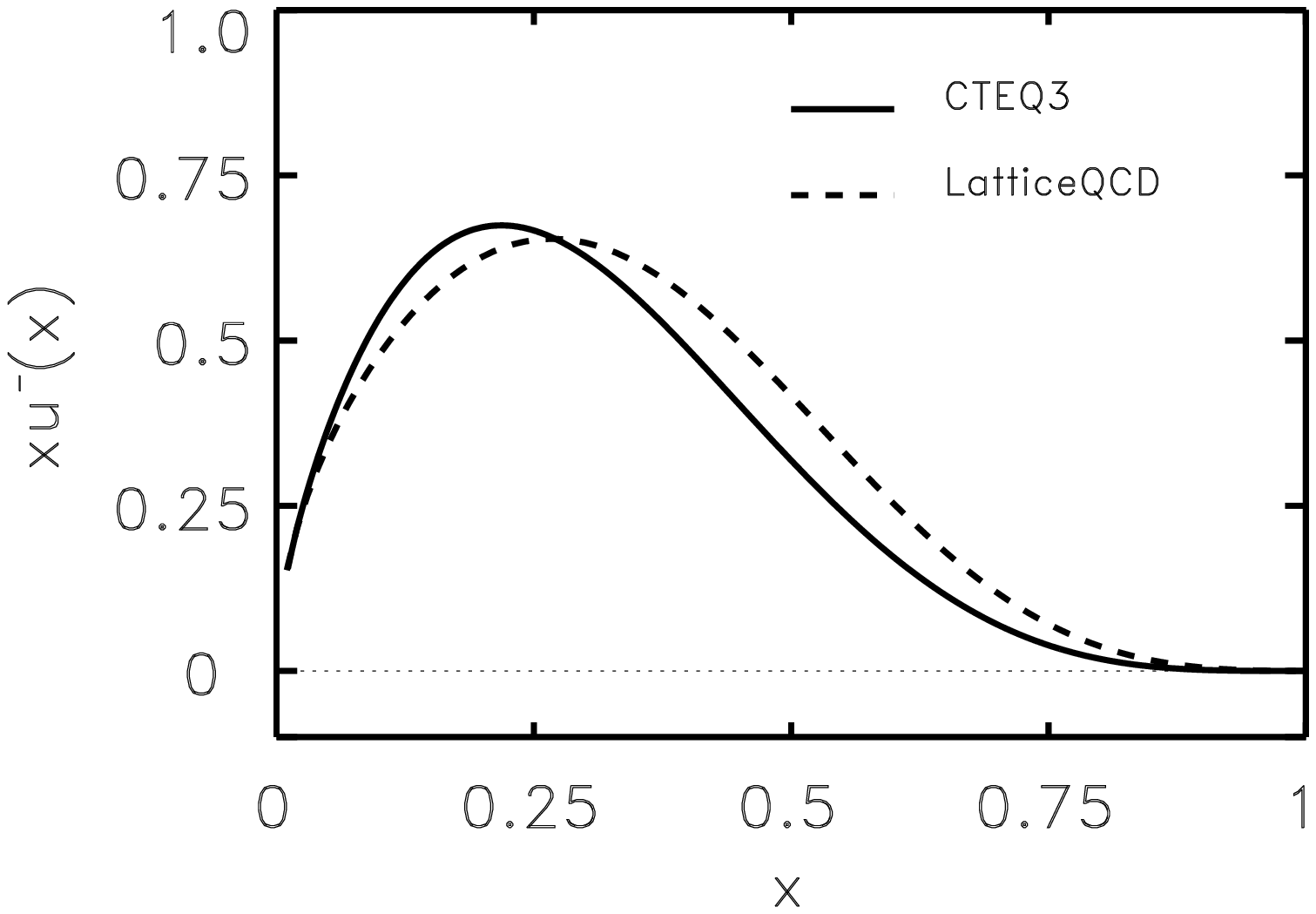}
\begin{description}
  
\item[Fig.~5] Reconstruction of the valence u-quark distribution function using
  lattice QCD results for the first non-trivial moment and the experimental
  normalization at small $x$, dashed line.  The solid line is the CTEQ
  parametrization \cite{CTEQ} at $\mu^2 = 2.5$ GeV$^2$.

\end{description}

\epsfbox{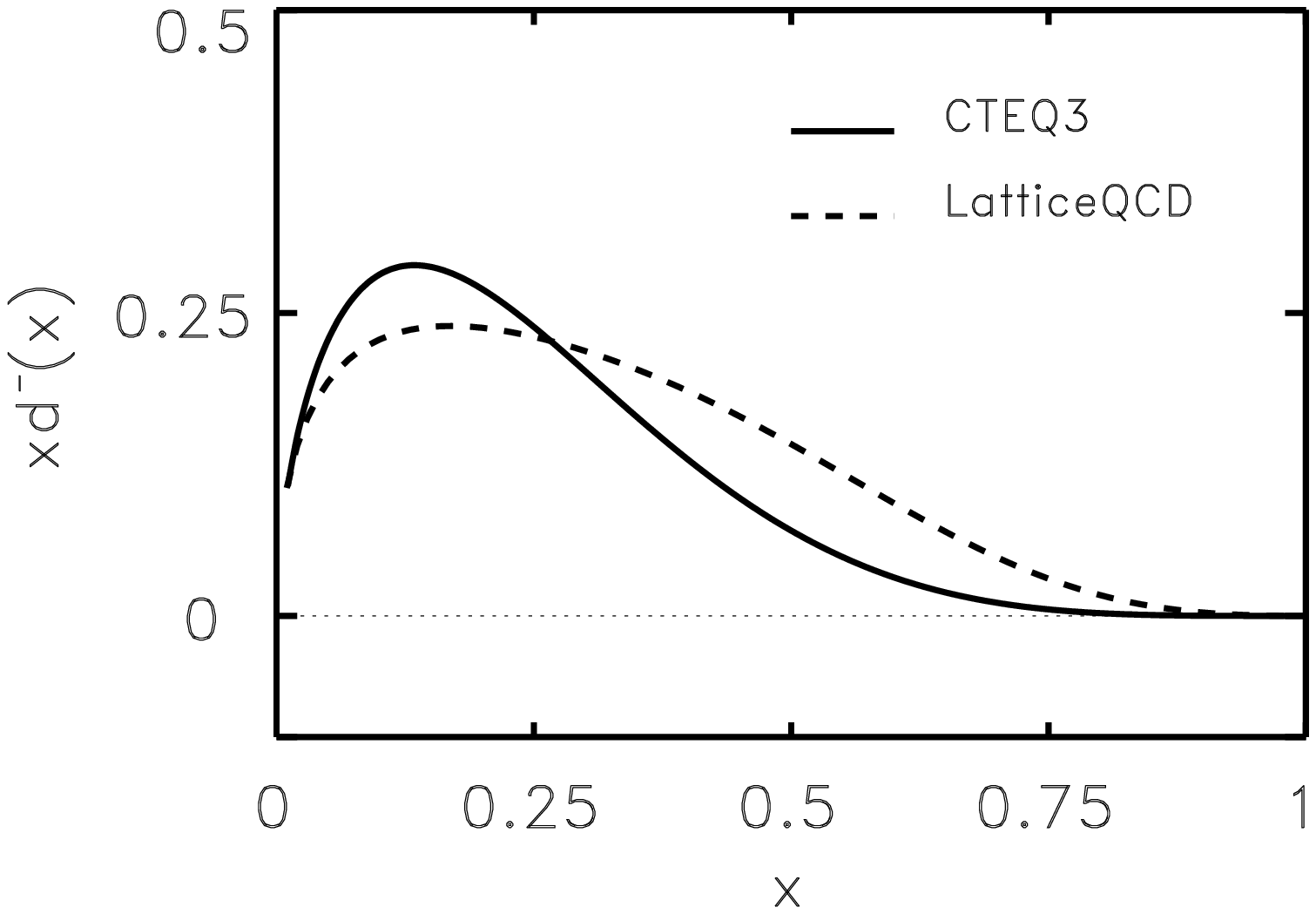}
\begin{description}
  
\item[Fig.~6] Reconstruction of the valence d-quark distribution function using
  the lattice QCD \cite{LAT95} results for the first non-trivial moment and
  the experimental normalization at small $x$. The solid line is the CTEQ
  parametrization \cite{CTEQ} at $\mu^2 = 2.5$ GeV$^2$.

\end{description}

\epsfbox{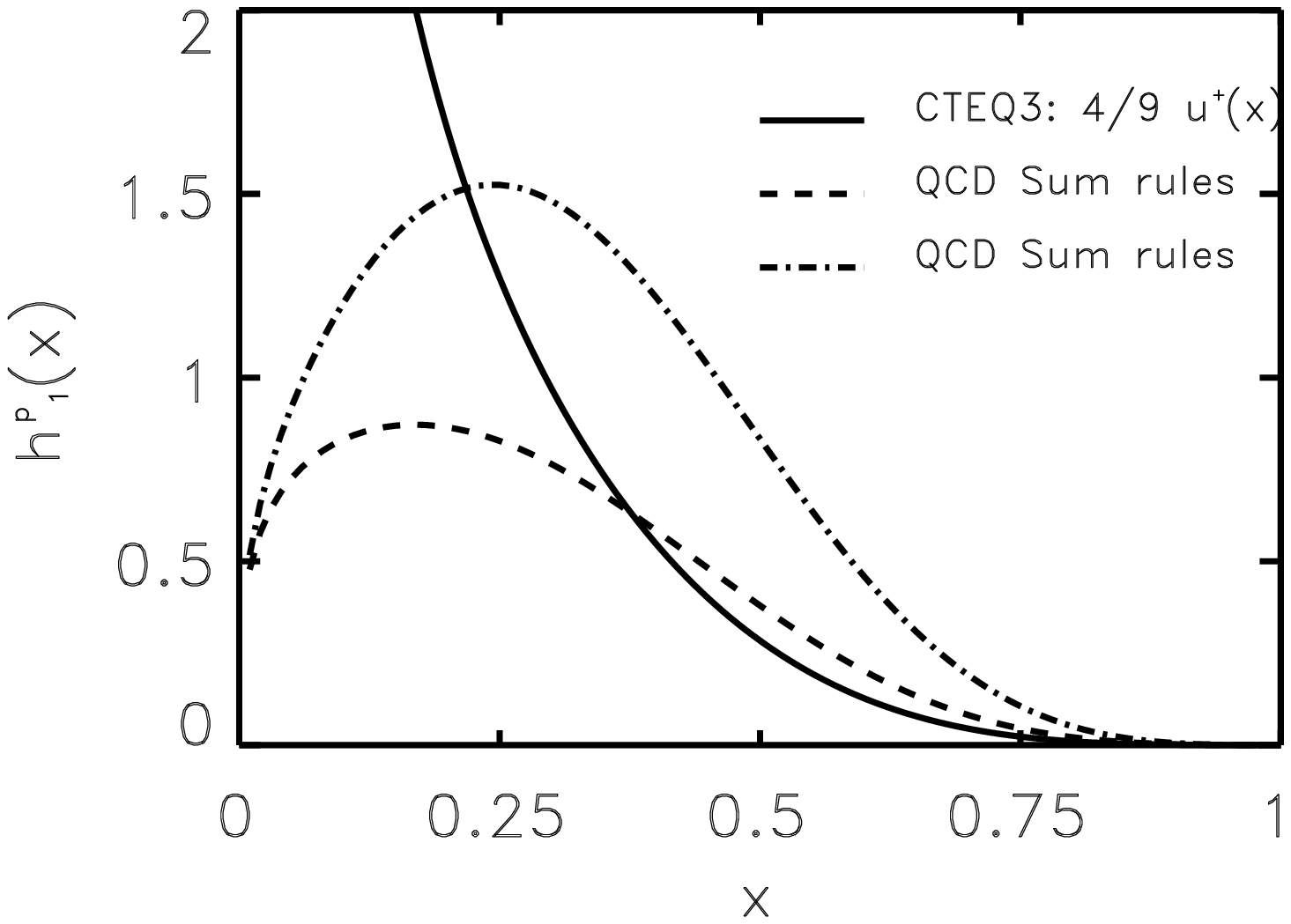}
\begin{description}
  
\item[Fig.~7] Reconstruction of the proton C-even transversity distribution
  $h_1^p(x)$, dashed and dot-dashed lines, using the QCD sum rules estimates
  for its first moment \cite{Ioff94}, and the asymptotics at large $z$ fixed by
  Regge arguments and comparison with the $g_1^p(x)$ data. The unpolarized
  u-quark distribution $\frac{4}{9}\, u(x,\mu^2)$ according to the CTEQ
  parametrization \cite{CTEQ} at $\mu^2$ = 2.5 GeV$^2$ is depicted by the
  solid line.    

\end{description}

\end{document}